\documentclass[12pt,fleqn]{article}

\usepackage{amssymb}
\usepackage{epsfig}
\usepackage{eucal}
\usepackage{amsmath}

\numberwithin{equation}{section}

\newlength{\dinwidth}
\newlength{\dinmargin}
\newcommand{\ab}{\bar{\al}_s}
\newcommand{\al}{\alpha}
\newcommand{\as}{\alpha_s}                   

\newcommand{\camint}{_{{1\over2}-\ui\infty}^{{1\over2}+\ui\infty}}

\newcommand{\de}{\partial}
\newcommand{\dif}{{\rm d}}

\newcommand{\De}{\Delta}

\newcommand{\esp}[1]{{\rm e}^{#1}}

\newcommand{\F}{{\CMcal F}}
\newcommand{\ga}{\gamma}
\newcommand{\gb}{\bar{\gamma}}
\newcommand{\G}{{\CMcal G}}                    

\newcommand{\kk}{{\boldsymbol k}}

\newcommand{\La}{\Lambda}
\newcommand{\om}{\omega}
\newcommand{\ord}{{\CMcal O}}                
\newcommand{\op}{\om_{\mathbb{P}}}           

\newcommand{\qq}{{\boldsymbol q}}

\newcommand{\si}{\sigma}
\newcommand{\tra}{-\hspace{-8.4pt}\raisebox{.5pt}{:}\hspace{5pt}}
\newcommand{\ts}{\textstyle}
\newcommand{\ui}{{\rm i}}

\begin{document}

\title{\large\bf
 The BFKL Equation at Next-to-Leading Level and Beyond
   \footnote{Work supported in part by the E.U. QCDNET contract
            FMRX-CT98-0194 and by MURST (Italy).}}
\author{M. Ciafaloni$^{\dagger}$ and D. Colferai$^{\dagger}$\\
{\em Dipartimento di Fisica, Universit\`a di Firenze} \\
{\em and INFN, Sezione di Firenze} \\
{\em Largo E. Fermi, 2 - 50125  Firenze}}
\date{}
\maketitle
\thispagestyle{empty}
\begin{abstract}
On the basis of a renormalization group analysis of the kernel and of
the solutions of the BFKL equation with subleading corrections, we
propose and calculate a novel expansion of a properly defined
effective eigenvalue function. We argue that in this formulation the
collinear properties of the kernel are taken into account to all
orders, and that the ensuing next-to-leading truncation provides a
much more stable estimate of hard Pomeron and of resummed
anomalous dimensions.
\end{abstract}
\begin{center}
\end{center}
\vspace*{2.5 cm}
{\small $~^{\dagger}$ e-mail: ciafaloni@fi.infn.it, colferai@fi.infn.it}
\newpage


The recent calculation \cite{cc96,flcc98} of the next-to-leading
$\log s$ (NL) corrections to the kernel of the BFKL equation has solved a
long-standing problem \cite{fl89}, but has also raised some puzzling
questions.
In fact, such corrections -- which decrease both the gluon anomalous
dimension and the hard pomeron intercept -- turn out to be so large,
that they cannot be taken literally for reasonable values of
$\as\geq0.1$. The question then arises: what is the origin of such
large corrections? and, what use can we make of them?

We do not suggest here to perform exact higher order calculations, not 
only because they are awkward, but also because they mix with
unitarity effects \cite{lmb} and thus cannot be described
within the BFKL equation 
alone. Our purpose is rather to illustrate the physics of
subleading corrections and to perform {\em partial resummations} on
the basis of a renormalization group analysis which improves the
perturbative formulation of the small-$x$ problem.

Our argument follows two steps. First, we propose an ansatz for the
BFKL solutions which is automatically consistent with the R.G. for a
general dependence of the kernel on the running coupling. Secondly, we 
identify the subleading corrections to be resummed, mainly the
scale-dependent ones, as recently suggested \cite{s98,ctalksp},
and the collinear
singular ones \cite{cc97}. The improved NL expansion follows naturally.


\section{Renormalization Group Analysis}

In small-$x$ physics ($s\gg Q^2\gg \La^2$), Regge theory and the
renormalization group come to a sort of clash, which provides
nontrivial consistency requirements.

It is known \cite{ck89,ccpom97} that the BFKL equation
satisfies R.G. factorization in
an asymptotic way. Introducing, by $\kk$-factorization \cite{cch90}, the NL
gluon Green's function
\begin{equation}
 \G_{\om}(t,t_0)=\big[\om-\ab(t)\big(K_0+\ab(\mu^2)\,K_1\big)\big]^{-1}
 \qquad,\qquad\left(\ab={N_c\as\over\pi}\right)\quad,\label{defg}
\end{equation}
its asymptotic form for $t\equiv\log\kk^2/\La^2\gg t_0$ is given by
\begin{equation}
 \G_{\om}(t,t_0)\simeq C(\om,\as(t))\left[\exp\int_{t_0}^{t}
 \ga_{\om}^+(\as(\tau))\dif\tau\right]C_0(\om,\as(t_0))\quad,\label{gasint}
\end{equation}
where $\ga_{\om}^+$ is the larger eigenvalue of the singlet anomalous
dimension matrix, defined by the equation
\begin{equation}
 \om=\ab(t)\big(\chi_0(\ga_{\om}^+)+\ab(\mu^2)\chi_1(\ga_{\om}^+)\big)
 \quad.\label{defgamma}
\end{equation}
Solutions to this equation are found so long as
$\om\geq\op(\as)$, the ``hard pomeron'' intercept, which in
this treatment is estimated from eq.~(\ref{defgamma}) and from the NL 
calculations \cite{flcc98} to be
\begin{equation}
 \op(\as)=\ab\big(\chi_0(\ts{1\over2})+\ab\chi_1(\ts{1\over2})
 \big)\simeq2.77\ab(1-6.47\ab)\quad,\label{pomerone}
\end{equation}
leading to the anomalously large NL corrections mentioned before.

While the $t$-dependent coefficient $C(\om,\as(t))$ in eq.~(\ref{gasint}) is
perturbatively calculable, the $t_0$-dependent one $C_0$ is not,
because it contains the leading
$\om$-singularity, that we call simply the ``pomeron''.
 The latter turns out to be {\em dependent} on how
$\as(t)$ is smoothed out or cut-off around $t=0$ ($\kk^2=\La^2$), in
order to avoid the Landau pole. However if $t_0\gg1$, it is suppressed 
by a power of $\La^2/\kk_0^2$.

Perturbative calculations are thus hampered by two
$\om$-singularities. The first one, the hard pomeron $\op(\as(t))$ is
a singularity of the {\em anomalous dimension expansion}, not
necessarily of the full amplitude, and is our main concern here. It
dominates an intermediate-$x$ regime, characterized by quite large
anomalous dimensions.
The second singularity -- the pomeron $\op$ -- is the leading
$\om$-singularity  of the amplitude and dominates the very small-$x$
regime, but is non-perturbative and is not directly related to scaling 
violations.

The treatment of the hard pomeron just summarized has, however, the
limitation that $\as(\mu^2)$ in eq.~(\ref{defg}) is kept frozen in
front of $K_1$, while the overall factor $\as(t)$ is allowed to
run. Even if consistent at NL level, this approach should be improved
in order to perform partial resummations to all orders in $\as(t)$, as 
we envisage here. How can this be achieved?

Our main proposal is to use $\om$ as the expansion parameter of the
BFKL solutions, instead of $\as(t)$. More precisely, we look for
the (regular) solution \cite{ccpom97} of the homogeneous BFKL equation
\begin{equation}
 \om\F_{\om}(t)=[K_{\om}\F_{\om}](t)\quad,\label{eqbfkl}
\end{equation}
where the kernel $K_{\om}$ has a general $\as(t)$-dependence of the
form (see sec.~(\ref{intle}))
\begin{equation}
 K_{\om}(\kk,\kk^{'})=\ab(t)K^{\om}_0(\kk,\kk^{'})+
 \ab(t)^2K^{\om}_1(\kk,\kk^{'})+\cdots\qquad,\qquad
 t\equiv\log{\kk^2\over\La^2}\quad,\label{nucleo}
\end{equation}
and the $K^{\om}_n:n=0,1,\cdots$ are scale invariant kernels which may 
be $\om$-dependent.

We then assume the ansatz
\begin{equation}
 \F_{\om}(t)={1\over\kk^2}\int\camint{\dif\ga\over2\pi\ui}\,
 \esp{\ga t-{1\over b\om}X(\ga,\om)}\qquad,\qquad
 b\equiv{\pi\over N_c}{11N_c-2N_f\over12\pi}   \quad,\label{rappf}
\end{equation}
where $X(\ga,\om)$ is to be found by solving eq.~(\ref{eqbfkl}), as a
power series in $\om$.
The relevance of such solution is
that it is shown to represent \cite{ccs99} the asymptotic form of the Green's
function (\ref{defg}) for a general form of the kernel $K$, as follows
\begin{equation}
 \G_{\om}(t,t_0)\simeq\F_{\om}(t)\tilde{\F}_{\om}(t_0)\qquad,
 \qquad(t\gg t_0\gtrsim1)\label{fattg}
\end{equation}
where, for a given regularization of the Landau pole, $\F_{\om}(t)$
has the form
(\ref{rappf}), while $\tilde{\F}_{\om}(t_0)$ contains the 
(regularization-dependent) pomeron singularity.

While the  detailed properties of $X(\ga,\om)$ are dependent on how
$\as(t)$ in eq.~(\ref{nucleo}) is smoothed out or cut-off around $t=0$, 
the large-$t$ behaviour of eq.~(\ref{rappf}) is instead universal,
provided a saddle point $\gb_{\om}(t)$ of the exponent
$E_{\om}(\ga,t)$ of eq.~(\ref{rappf}) exists and is stable. We thus
assume, in the asymptotic regime $bt\gtrsim{1\over\om}\gg1$, the
expansion
\begin{equation}
 E_{\om}(\ga,t)=E_{\om}(\gb_{\om},t)-{1\over2b\om}\,\chi^{'}(\gb_{\om},\om)
 (\ga-\gb_{\om})^2-{1\over6b\om}\,\chi^{''}(\gb_{\om},\om)
 (\ga-\gb_{\om})^3+\cdots\quad,\label{espans}
\end{equation}
with the saddle point conditions
\begin{equation}
 b\om t\equiv\chi(\gb_{\om}(t),\om)=X^{'}(\gb_{\om}(t),\om)\quad,
 \quad\chi^{'}(\gb_{\om}(t),\om)<0\qquad,\qquad
 (\#)^{'}\equiv{\de\over\de\ga}(\#)\quad,\label{cond}
\end{equation}
which need to be checked ``a posteriori''.

The crucial property of the representation (\ref{rappf}) around the
saddle point of eqs.~(\ref{espans}) and (\ref{cond}) is that it takes 
a form consistent with the R.G., namely
\begin{equation}
 \kk^2\F_{\om}(t)\sim{1\over\sqrt{2\pi\big(-\chi^{'}(\gb_{\om}(t),\om)\big)}}
 \,\exp\big\{E_{\om}(\gb_{\om}(t),t)\big\}\times
 \big(1+\ord(\om)\big)\quad,\label{rg}
\end{equation}
where, for any $\om$-dependence,
\begin{equation}
 E_{\om}(\gb_{\om}(t),t)=\gb_{\om}(t)\,t-{1\over b\om}\,X(\gb_{\om}(t),\om)=
 \int^t \gb_{\om}(\tau)\dif\tau+\text{const}\quad,\label{rgesp}
\end{equation}
so that eq.~(\ref{fattg}) takes the form anticipated in
eq.~(\ref{gasint}) for the gluon Green's function.

Furthermore, the form of $\chi(\ga,\om)$ can be found, as an expansion 
in $\om$, from the original equation (\ref{eqbfkl}), expanded around
the saddle point. For instance, if we let the coupling run up to NL
level, by expanding the eigenvalue functions
$\chi_0^{\om}(\ga),\chi_1^{\om}(\ga),\cdots$ of
$K_0^{\om},K_1^{\om},\cdots$
up to the relevant order in $\ga-\gb$, we find  the solution
\begin{equation}\hspace{-14pt}
 \chi(\ga,\om)=\chi_0^{\om}(\ga)+\om\,{\chi_1^{\om}(\ga)
 \over\chi_0^{\om}(\ga)}+\om^2\,{1\over\chi_0^{\om}(\ga)}\left(
 {\chi_2^{\om}(\ga)\over\chi_0^{\om}(\ga)}-\left({\chi_1^{\om}(\ga)
 \over\chi_0^{\om}(\ga)}\right)^2+b\left({\chi_1^{\om}(\ga)\over
 \chi_0^{\om}(\ga)}\right)^{'}\right)+\ord(\om^3)\label{svilchi}
\end{equation}
which has been pushed up to NNL level in $\om$. Some of the terms of the
$\om$-expansion (\ref{svilchi}) follow from the trivial replacement
${1/t}=b\om/\chi_0^{\om}(\ga)+\cdots$,
while the term $b\big(\chi_1^{\om}(\ga)/\chi_0^{\om}(\ga)\big)^{'}$
needs a careful treatment of fluctuations. This result is confirmed, and
extended to higher orders, by the replacement $t\to\de_{\ga}$ in the 
BFKL equation, which leads to a non linear differential equation
for $\chi$ \cite{ccs99}.

The result (\ref{svilchi}) is the basic formula that we present here, in
order to resum some large subleading corrections, and to obtain a
smoother NL truncation for the remaining ones. Basically, the
coefficients of the $\om$-expansion (\ref{svilchi}) turn out to be less
singular than the ones in the $\as$-expansion in the collinear limits 
$\ga\to0$ ($\ga\to1$), related to $t\gg t_0$ ($t_0\gg t$).
This is due to the
powers of $\chi_0$ in the denominators which damp the collinear
behaviour and to cancellations to be explained below. This feature
will yield a smoother expansion in the region $\ga\simeq1/2$ also,
which is the relevant one for the hard pomeron estimate.


\section{Improved next-to-leading expansion}\label{intle}

Before proceeding further, we need to understand the physics of
subleading corrections, in order to single out the ones that need to be
resummed. We limit ourselves, for simplicity, to the gluonic contributions,
because the $q\bar{q}$ ones turn out to be small \cite{cc96}.

The explicit eigenvalues of the gluonic part of the kernels $K_0$ and
$K_1$ are \cite{flcc98}
\begin{equation}
 \chi_0(\ga)=2\psi(1)-\psi(\ga)-\psi(1-\ga)
 \underset{\ga\to0}{\simeq}{1\over\ga}-2\psi(1)\,\ga^2+\cdots\label{chio}
\end{equation}
and
\begin{align}
 \chi_1(\ga)=&\left(-{11\over24}(\chi_0^2(\ga)+\chi_0^{'}(\ga))
 \right)+\left[-{1\over4}\,\chi_0^{''}(\ga)\right]+\nonumber\\
 +&{1\over4}\left\{\!\left({67\over9}-{\pi^2\over3}\right)\!\chi_0(\ga)-
 \left({\pi\over\sin\pi\ga}\right)^2\!\!{\cos\pi\ga\over3(1-2\ga)}\left(11+
 {\ga(1-\ga)\over(1+2\ga)(3-2\ga)}\right)\right\}+\nonumber\\
 +&{3\over2}\,\zeta(3)+{\pi^3\over4\sin\pi\ga}-\Phi(\ga)
 \quad,\label{chiu}\\
 \Phi(\ga)\equiv&\sum_{n=0}^{\infty}(-)^n\left[{\psi(n+1+\ga)-\psi(1)\over
 (n+\ga)^2}+{\psi(n+2-\ga)-\psi(1)\over(n+1-\ga)^2}\right]\quad.\nonumber
\end{align}
Here we quote the form of $\chi_1$ referring to the choice
$s_0=k k_0$ of the energy scale in the NL $\kk$-factorization formula 
\cite{cch90}:
\begin{equation}
 {\dif\si^{AB}\over\dif^2\kk\,\dif^2\kk_0}=\int{\dif\om\over2\pi\ui}\;
 h_A(\kk)\,h_B(\kk_0)\,\G_{\om}(\kk,\kk_0)
 \left({s\over k k_0}\right)^{\om}\quad,\label{sezurto}
\end{equation}
where $\kk$ and $\kk_0$ are the transverse momenta of the outgoing
jets in the fragmentation regions of the incoming partons $A$ and $B$
respectively.

In the expansion (\ref{chiu}) we have singled out some contributions
which have a natural physical interpretation, namely the running
coupling terms (in round brackets), the scale-dependent terms (in
square brackets) and the collinear terms (in curly brackets).

The running coupling terms are proportional to the one-loop beta
function coefficient $b$ and refer to the choice of eq.~(\ref{nucleo}) of 
factorizing the running coupling $\as(t)$ at the upper scale $\kk^2$
of the kernel $K_{\om}(\kk,\kk^{'})$,
instead of a symmetrical combination of $\kk$ and $\kk^{'}$. Their explicit
form can be shown \cite{ctalksp} to shift the effective scale to the symmetrical
one $\qq^2=(\kk-\kk^{'})^2$, which is the one occurring in the phase
space $\nu\gtrsim\qq^2$, where $\nu$ is the longitudinal part of the
two-gluon subenergy in
the $s$-channel. In $\ga$-space, the running coupling terms show a
double pole at $\ga=1$ only.

The collinear and scale-dependent terms have multiple poles at both
$\ga=0$ and $\ga=1$ due to the behaviour of the kernel for $k\gg k^{'}$
and $k^{'}\gg k$. All these poles are actually of collinear origin, and
the cubic ones are dependent on the choice of the scale of the energy in
eq.~(\ref{sezurto}), whether it is $s_0=kk_0$ (the symmetrical
choice), or instead $s_1=k^2$ ($s_2=k_0^2$), which is the choice 
leading to the correct scaling variable $k^2/s$ ($k_0^{2}/s$) for
$k\gg k_0$ ($k_0\gg k$).

It has been already remarked \cite{c98} that the choice $s_0=kk_0$ is not
collinear safe and leads to the nasty singularities of type $1/\ga^3$
($1/(1-\ga)^3$) of $\chi_1$ in eq.~(\ref{chiu}), which correspond to
the double logarithmic behaviour $\sim\log^2k/k^{'}$ of the kernel
when $k\gg k^{'}$ ($k^{'}\gg k$).

On the other hand, complete information about the collinear
behaviour of the kernel in such regions comes from the renormalization 
group equations. In general the kernel $K_{\om}(t,t^{'})$ in
eq.~(\ref{nucleo}) is collinear finite and symmetrical in its
arguments for $s_0=kk_0$, so that it must have the form
\begin{equation}
 K_{\om}(\as(\mu^2),\mu^2,\kk,\kk^{'})=
 {\ab(t)\over\kk^2}\,\hat{K}_{\om}(\as(t);t,t^{'})=
 {\ab(t^{'})\over\kk^{'2}}\,\hat{K}_{\om}(\as(t^{'});t^{'},t)
 \quad,\label{rgeq}
\end{equation}
where $\hat{K}$ can be expanded in $\as(t)$ with scale-invariant
coefficients, as assumed in eq.~(\ref{nucleo}).

Furthermore, calling $K_{\om}^{(1)}$ ($K_{\om}^{(2)}$) the kernel at
energy scale $k^2$ ($k_0^{2}$), it acquires for $k^{'}/k\to0$
($k/k^{'}\to0$) the collinear singularities due to the non singular
part of the gluon anomalous dimension in the $Q_0$-scheme \cite{c95} which,
neglecting the $q\bar{q}$ part, is
\begin{align}
 \tilde{\ga}(\om)&=\ga_{gg}(\om)-{\ab\over\om}=\ab A_1(\om)+
 \ab^2A_2(\om)+\cdots\quad,\label{dimanom}\\
 A_1(\om)&=-{11\over12}+\ord(\om)
 \quad,\quad A_2(\om)=0+\ord(\om)\quad,\nonumber
\end{align}
the singular part being taken into account by the BFKL iteration itself.
It follows that, for $k\gg k^{'}$,
\begin{align}
 K_{\om}^{(1)}(\as(t);t,t^{'})&\simeq{\ab(t)\over\kk^2}\;\exp\int_{t^{'}}^t
 \tilde{\ga}(\om,\as(\tau))\;\dif\tau\nonumber\\
 &\simeq{\ab(t)\over\kk^2}\left(1-b\ab(t)\log{\kk^2\over\kk^{'2}}\right)^
 {-{A_1(\om)\over b}}\quad,\label{kuno}
\end{align}
with a similar behaviour, with  $t$ and $t^{'}$ interchanged, for
$K_{\om}^{(2)}$ in the opposite limit $k^{'}\gg k$.

In order to derive from eq.~(\ref{kuno}) the $\ga$-dependent
singularities of $K_{\om}$ care should be taken of the relationships \cite{c98}
\begin{equation}
 K_{\om}(\kk,\kk^{'})=\left({k^{'}\over k}\right)^{\om}K_{\om}^{(1)}
 (\kk,\kk^{'})=\left({k\over k^{'}}\right)^{\om}K_{\om}^{(2)}
 (\kk,\kk^{'})\label{komega}
\end{equation}
which shift the $\ga$-singularities of $K_{\om}^{(1)}$ ($K_{\om}^{(2)}$)
by $-\om/2$ ($+\om/2$). As a consequence, by eq.~(\ref{kuno}), the
$\ga$-singularities of the eigenvalue functions $\chi^{\om}_n$ of the 
kernels $K^{\om}_n$ in eq.~(\ref{nucleo}) are
\begin{align}
 \chi^{\om}_n(\ga)&\simeq{1\cdot A_1(A_1+b)\cdots(A_1+(n-1)b)\over
 (\ga+{1\over2}\om)^{n+1}}\quad,\quad(\ga\ll1)\nonumber\\
 &\simeq{1\cdot(A_1-b)(A_1-2b)\cdots(A_1-nb)\over
 (1-\ga+{1\over2}\om)^{n+1}}\quad,\quad(1-\ga\ll1)\quad.\label{chin}
\end{align}
In particular, $\chi_0^{\om}$ has symmetrical singularities and
$\chi^{\om}_1$ slightly asymmetrical ones, as follows:
\begin{align}
 \chi^{\om}_0(\ga)&\simeq{1\over\ga+{1\over2}\om}+{1\over1-\ga+
 {1\over2}\om}\quad,\label{chiosv}\\
 \chi^{\om}_1(\ga)&\simeq{A_1\over(\ga+{1\over2}\om)^2}+{A_1-b
 \over(1-\ga+{1\over2}\om)^2}\quad.\label{chiusv}
\end{align}

The shift by $\pm\om/2$ in eqs.~(\ref{chin})-(\ref{chiusv}) already
represents a resummation of the scale-dependent singularities, as
noticed by Salam \cite{s98}, and it provides the correct scale-dependent
terms ($\sim1/\ga^3\;,\;1/(1-\ga)^3$) by the
$\om$-expansion of eq.~(\ref{chiosv}). On the other hand, eq.
(\ref{chiusv}) provides the correct collinear behaviour of $K_{\om}^{(1)}$
($K_{\om}^{(2)}$) close to $\ga=0$ ($\ga=1$) at NL order, and eq.
(\ref{chin}) generalizes it to all orders.

The simplest realization of eq.~(\ref{chiosv}) is to assume, following
ref.~\cite{s98}, that $\chi^{\om}_0$ in eqs.~(\ref{nucleo}) and (\ref{svilchi}) is
provided by the eigenvalue function of the Lund model \cite{ags96}
\begin{equation}
 \chi^{\om}_0(\ga)=\psi(1)-\psi(\ga+\ts{1\over2}\om)+\psi(1)-
 \psi(1-\ga+\ts{1\over2}\om)\quad,\label{lund}
\end{equation}
which corresponds to the kernel
\begin{equation}
 K_0^{\om}(\kk,\kk^{'})=K_0(\kk,\kk^{'})\left({k_<\over k_>}
 \right)^{\om}\qquad,\qquad k_<\equiv\min\{k,k^{'}\}\quad,
 \quad k_>\equiv\max\{k,k^{'}\}\quad.\label{klund}
\end{equation}
Here the $\om$-dependent factor $(k_</k_>)^{\om}$ can be directly
interpreted as an $s$-channel threshold factor \cite{ctalkd}
in properly defined Toller  variables \cite{t65}.

The NL terms in eq.~(\ref{svilchi}) are now simply identified from
eq.~(\ref{chiu}) after subtraction of the singular part of the 
scale-dependent terms already taken into account in eq.~(\ref{lund}).
This subtraction eliminates the cubic singularities in $\chi_1$, but leaves
double and single poles which need to be shifted by the procedure of
ref.~\cite{s98}. This amounts to the replacement of $\chi_1$ by
\begin{align}
 \chi_1^{\om}(\ga)\equiv\,&\chi_1(\ga)+{1\over2}\chi_0(\ga){\pi^2\over
 \sin^2(\pi\ga)}+{\pi^2\over6}\,(\chi_0^{\om}(\ga)-\chi_0(\ga))+
 A_1(\om)\psi^{'}(\ga+{\ts{1\over2}}\om)\nonumber\\
 &-A_1(0)\psi^{'}(\ga)+
 (A_1(\om)-b)\psi^{'}(1-\ga+{\ts{1\over2}}\om)-
 (A_1(0)-b)\psi^{'}(1-\ga)\nonumber\\
 \equiv\,&A_1(\om)\psi^{'}(\ga+{\ts{1\over2}}\om)+(A_1(\om)-b)\psi^{'}(1-\ga+
 {\ts{1\over2}}\om)+{\pi^2\over6}\,\chi_0^{\om}(\ga)+\tilde{\chi}_1(\ga)
 \quad,\label{rim}
\end {align}
where now $\tilde{\chi}_1(\ga)$ has no $\ga=0$ nor $\ga=1$ singularities 
at all.

What about higher subleading orders in eq.~(\ref{svilchi})? Here the
crucial observation is that, after substitution of the behaviour
(\ref{chin}) in the NNL term of eq.~(\ref{svilchi}), the leading
$\ga$-singularities {\em cancel out close to both $\ga=0$ and}
$\ga=1$. Therefore, the coefficient of $\om^2$ in eq.~(\ref{svilchi})
has {\em no $\ga$-singularities} at all, and no further resummation
is needed. This peculiar feature, confirmed at higher orders \cite{ccs99},
is due to the fact that running
coupling effects are already taken into account by the representation
(\ref{rappf}) with the solution (\ref{svilchi}).

\begin{figure}[t]
 \centering
\begin{picture}(120,72)
  \put(0,-7){\includegraphics[width=120mm]{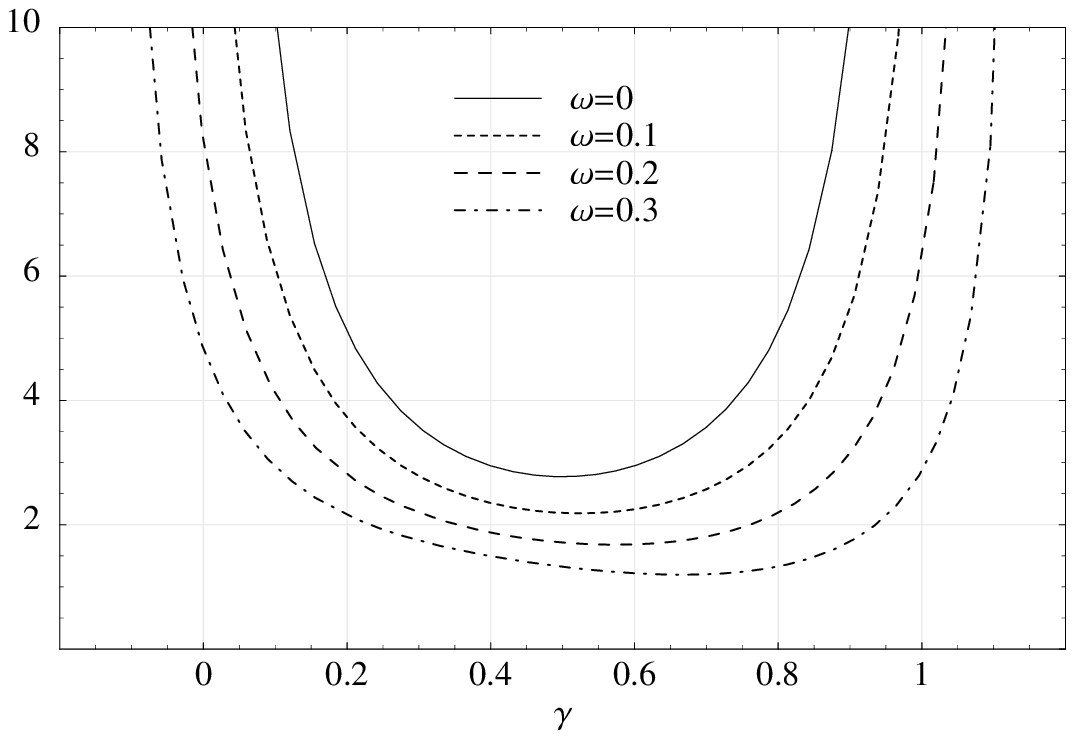}}
  \put(-10,36){\small$\chi(\ga,\om)$}
\end{picture}
  \caption{\small\sl Resummed eigenvalue function $\chi(\ga,\om)$ for various
 values of $\om$.}
\label{figchi}
\end{figure}

Our truncated NL effective eigenvalue function (fig.~\ref{figchi}) 
then reads
\begin{equation}
 \chi(\ga,\om)=\chi_0^{\om}(\ga)+\om\,{\chi_1^{\om}(\ga)\over
 \chi_0^{\om}(\ga)}+\ord(\om^2)\quad,\label{chieff}
\end{equation}
and contains only the shifted single poles in $\ga$,
while the neglected terms have no leading twist $\ga$-singularities.
This expression performs the resummation of the scale-dependent
singularities proposed in  ref.~\cite{s98}, but it also takes into account the
collinear singularities of eq.~(\ref{chin}) to all orders.
This resummation effect is perhaps more easily seen by using
eq.~(\ref{chieff}) in order to recast eq.~(\ref{cond}) in the form
\begin{equation}
 b\om t=\chi_0^{\om}\left[1-{1\over bt}\left({\chi_1^{\om}\over
 \chi_0^{\om}}+\ord(\om)\right)\right]^{-1}\quad,\label{bot}
\end{equation}
which defines an effective $\chi$-function as a power series in $\as$.
By specializing eq.~(\ref{bot}) to the scale $\kk^2$ (shift $\ga+\om/2\to
\ga$), it is apparent that all the small $\ga$ poles $(A_1(\om)/\ga bt)^n$
are resummed as a geometric series, to provide the full one-loop
anomalous dimension of eq.~(\ref{dimanom}).

The expression (\ref{chieff}) for $\chi(\ga,\om)$ is not fully symmetrical
under the replacement $\ga\leftrightarrow(1-\ga)$, but keeps the
asymmetrical $-b\om\chi^{'}_0/2\chi_0$ contribution of eq.~(\ref{chiu}).
The breakdown of the $\ga\leftrightarrow(1-\ga)$ symmetry in
eq.~(\ref{rappf}) follows from that of scale invariance by running
coupling effects. It is easy to recognize that the contribution
$-b\om\chi^{'}_0/2\chi_0$ to $\chi(\ga,\om)$ yields a factor
$\sqrt{\chi_0^{\om}(\ga)}$ in the representation (\ref{rappf}), and this
factor {\em is required} \cite{ccs99} by the continuum normalization of the
eigenfunctions of $K_{\om}$ provided by the Jacobian
$X^{'}(\ga,\om)=\chi_0^{\om}(\ga)+\ord(\om)$.
Therefore, even if the kernel $K_{\om}$ and the Green's function
$\G_{\om}$ are
symmetrical operators, the effective eigenvalue function $\chi(\ga,\om)$
must have some asymmetry, due to running coupling effects.

It appears from fig.~(\ref{figchi}) that $\chi(\ga,\om)$, with the
$b$-dependent asymmetry pointed out before, is a decreasing function of
$\om$ because of the negative sign of NL corrections, and contains a stable
minimum for all the relevant $\om$ values. This means that in the present
approach the hard pomeron can be estimated up to sizeable values of
$\as=0.2\tra0.3$, and it turns out to be $\op=0.26\tra0.32$
in this range, which is a reasonable value for the HERA data
\cite{a96} at intermediate values or $Q^2$.

Since in the expansion of eq.~(\ref{chieff}) the neglected terms -- of order
$\om^2$ or higher -- have no $\ga$-singularities, they only affect the
evaluation of eq.~(\ref{cond}) by a roughly $\ga$-independent uncertainty
which corresponds to a change of scale $\De(bt)=\ord(\om)$, or
$\De\as=\ord(\om)\as^2$. This means that the error affecting the NL
truncation is uniformly of NNL order for all values of $\ga$ --
whether  $\ord(\as)$, $\ord(\as/\om)$ or $\ord(1)$ -- and therefore it
cannot change too much the $\op$-estimate given above.

A detailed analysis of the anomalous dimensions and of the hard pomeron
properties in the present approach is left to a parallel investigation \cite{ccs99}.

\subsection*{Acknowledgements}

We wish to thank Gavin Salam for interesting discussions and suggestions.


\end{document}